\def\bey{\begin{eqnarray}}
\def\eey{\end{eqnarray}}
\def\be{\begin{equation}}
\def\ee{\end{equation}}
\def\ba{\begin{array}}
\def\ea{\end{array}}

\documentclass{elsart}
\usepackage{amsmath,bm}
\usepackage{graphicx}
\usepackage{dcolumn}

\usepackage[normalem]{ulem}  
\usepackage[dvips]{color} 

\begin{document}

\begin{frontmatter}
\title{Symmetry energy and neutron star properties in the saturated Nambu-Jona-Lasinio model }
\author{ Si-Na Wei, Wei-Zhou Jiang, Rong-Yao Yang, Dong-Rui Zhang}
\address{Department of Physics, Southeast University,
Nanjing 211189, China}
\begin{abstract} In this work, we adopt the Nambu-Jona-Lasinio (NJL) model that ensures the nuclear matter saturation properties to study the density dependence of the symmetry energy. With the interactions constrained by the chiral symmetry, the symmetry energy shows  novel characters different from  those in  conventional mean-field models. First, the negative symmetry energy at high densities that is absent in relativistic mean-field (RMF) models can be obtained  in the RMF approximation by introducing a chiral isovector-vector interaction, although it would be ruled out by  the neutron star (NS) stability. Second,  with the inclusion of the isovector-scalar interaction  the symmetry energy  exhibits  a general softening  at high densities even for the large slope parameter of the symmetry energy.    The  NS properties obtained in the present NJL model can be in accord with the observations.     The NS maximum mass obtained  with various isovector-scalar couplings and momentum cutoffs is well above the $2M_\odot$, and  the NS radius obtained  well meets the limits extracted from recent measurements. In particular, the significant reduction of the canonical NS radius occurs with the moderate decrease of the slope of the symmetry energy.
\end{abstract}

\begin{keyword}
Symmetry energy, neutron stars, NJL model, \PACS  11.30.Rd \sep 21.65.Jk\sep 26.60.Kp \end{keyword}

\end{frontmatter}

\section{INTRODUCTION}

The nuclear symmetry energy is important for understanding  the reaction dynamics of heavy-ion collisions, the structures of  neutron- and proton-rich nuclei, and properties of neutron stars (NS) ~\cite{ba1,ba2,lat01}.
Though the symmetry energy, which is the energy difference per nucleon between pure neutron matter and symmetric matter, is well constrained at saturation density  to date~\cite{ba3,ch14,ts10,la13,wang15}, the density dependence of the symmetry energy is still poorly known especially at supra-normal
densities~\cite{ba2,zhen4}. The symmetry energy predicted by different models is rather diverse at high densities~\cite{yong5,wen6,br00,chen05,fu06,ho01,ji07a,lie7}. Unfortunately, the symmetry energy extracted from the data with various isospin diffusion models also suffers from the large uncertainty which diversifies in super-soft~\cite{xiao16},  soft~\cite{ru11}, and stiff~\cite{zhao17} forms at high densities. We note that new experiments to probe the high-density symmetry energy are also on the way~\cite{xiao14}. While different high-density behaviors of the symmetry energy are usually classified by the magnitude of the slope of the symmetry energy at saturation density, we may raise the question: Are there  new high-density behaviors  of the symmetry energy that can't be simply elaborated by the slope parameter.

On the other hand, the super-soft symmetry energy which  reaches the maximum and then turns to negative values at high densities can be obtained from some nonrelativistic models~\cite{br00,chen05}, while it can not be produced in the relativistic mean field (RMF) models~\cite{fu06,ho01,ji07a,lie7}.  For instance, the nonlinear RMF models~\cite{ho01}, the density-dependent  RMF models ~\cite{ji07a,le1}, and the point coupling  RMF models~\cite{ni1,ni2,p1} predict similar tendencies of symmetry energy, and no super-soft symmetry energy arises in these models~\cite{lie7}. Since the success of RMF models in interpreting the pseudospin symmetry~\cite{gi99,li15,ya14} and analyzing polarization observables in proton-nuclei reactions~\cite{jr8,bc9} indicates that the relativistic dynamics that includes  the large attractive scalar and repulsive vector~\cite{Wal74,Chin77,Ser86,Ring96,Ser97} is of special importance, we may ask whether the super-soft symmetry energy is incompatible with the relativistic covariance, or it is hidden in some special interactions that are not included in usual RMF models.

To answer these questions, let's first recall the prime importance of the chiral symmetry in the strong interaction.  In fact,   the chiral symmetry has served as a cornerstone to construct the effective QCD models of the strong interaction~\cite{winb79,winb91}. In the development of RMF models, the chiral symmetry has also played an important role in guiding the nonlinear form of  the meson self-interacting terms needed for the  appropriate in-medium effects~\cite{gell60,lee74,mo77,bog83,bog83n}.  To explore the novel high-density behaviors of the symmetry energy in the RMF approximation, it is  appropriate to adopt chiral models and thus constrain the relevant interactions with the chiral symmetry. Among  models  respecting the chiral symmetry in bulk matter~\cite{gell60,bog83,namb10,fn93,la11}, the  Nambu-Jona-Lasinio (NJL) model~\cite{namb10} and chiral-$\sigma$ model~\cite{gell60,bog83} are two popular ones. The NJL model was originally proposed to realize the
spontaneous symmetry breaking since the pion, as the Goldstone
boson, can be derived dynamically. With the quark degrees of freedom, the NJL model is considered as  an effective model for the QCD~\cite{kl92,vo91,bu05}. While it is not straightforward to construct the nucleons and describe nuclear matter due to the absence of the confinement in the NJL model~\cite{bent01},  it is economic to realize in the NJL model the spontaneous breaking of the chiral symmetry with nucleonic degrees of freedom~\cite{v11,in12,pa03,lee13}, like the chiral-$\sigma$ model.  In the hadron-level NJL model, the character of chiral symmetry is also  measured by the chiral condensate in the non-perturbative vacuum. In this work,  we thus study in the hadron-level NJL model the density dependence of the symmetry energy with the various interactions respecting the chiral symmetry.

Recently,  remarkable progresses in NS observations have been achieved. Accurate mass measurements  determined two large-mass NS's:   the radio pulsar J1614-2230 with mass of $M=1.97\pm0.04M_{\odot}$~\cite{pd14} and the J0348+0432 with mass of $M=2.01\pm0.04 M_{\odot}$~\cite{ja15}. However, there is no consensus on the extracted NS radius~\cite{jia15} reported in the literature~\cite{gu14,st16,bog13,pou14,hei14,latt14,oz15}, due to the systematic uncertainties involved in the distance measurements and theoretical analyses of the light spectrum~\cite{mil13,ha01,zh07,su11}.  In this work, we will then investigate whether the parametrizations of the present saturated NJL model can satisfy the NS mass constraint and provide some useful comparisons with various NS radius constraints. In the following, we will in turn present the formalism, analyze the results,  and give the summary.

\section{Formalism}
The original NJL model that only contains scalar, pseudoscalar, vector and axial vector interactions can not reproduce saturation properties of nuclear matter. In order to obtain the saturation property, the scalar-vector (SV) interaction, which also respects the chiral symmetry,  was introduced~\cite{v11,in12}. This is similar to the chiral-$\sigma$ model, where the saturation is fulfilled by introducing the scalar-vector coupling~\cite{bog83,uch10}. Similar efforts were also made  to study the nuclear matter saturation and the phase diagram in the NJL model~\cite{pa03,lee13}.  The Lagrangian of the saturated NJL model can then be written as ~\cite{in12}:
\begin{eqnarray} \label{eq1}
 \mathcal{L}_{0}&=&\bar{\psi}(i\gamma_{\mu}{\partial^{\mu}}-m_0)\psi+\frac{G_S}{2}[(\bar{\psi}\psi)^2- (\bar{\psi}\gamma_5\tau\psi)^2]
-\frac{G_V}{2}[(\bar{\psi}\gamma_{\mu}\psi)^2+(\bar{\psi}\gamma_{\mu}\gamma_5\psi)^2]\nonumber\\
&&+\frac{G_{SV}}{2}[(\bar{\psi}\psi)^2-(\bar{\psi}\gamma_5\tau\psi)^2] \cdot[(\bar{\psi}\gamma_{\mu}\psi)^2+ (\bar{\psi}\gamma_{\mu}\gamma_5\psi)^2],
\end{eqnarray}
where $m_0$ is the bare nucleon mass.  $G_S$, $G_V$ and $G_{SV}$ are the scalar, vector  and scalar-vector coupling constants, respectively. It is easy to see that the Lagrangian is chiral symmetric when $m_0=0$. In order to investigate the density dependence of the symmetry energy, we introduce the isovector,  isovector-vector and isovector-scalar interactions in the Lagrangian which are  written as:
\begin{eqnarray} \label{eq2}
 \mathcal{L}_{IV}&=&\frac{G_{\rho}}{2}[(\bar{\psi}\gamma_{\mu}\tau\psi)^2+ (\bar{\psi}\gamma_{\mu}\gamma_{5}\tau\psi)^2]
 + \frac{G_{\rho{V}}}{2}[(\bar{\psi}\gamma_{\mu}\tau\psi)^2+(\bar{\psi}\gamma_{\mu}\gamma_{5}\tau\psi)^2] \cdot\nonumber\\
&& [(\bar{\psi}\gamma_{\mu}\psi)^2+ (\bar{\psi}\gamma_{\mu}\gamma_5\psi)^2]
 +\frac{G_{\rho{S}}}{2}[(\bar{\psi}\gamma_{\mu}\tau\psi)^2+(\bar{\psi}\gamma_{\mu}\gamma_{5}\tau\psi)^2]\cdot\nonumber\\
&& [(\bar{\psi}\psi)^2- (\bar{\psi}\gamma_5\tau\psi)^2],
\end{eqnarray}
where  $G_{\rho}$, $G_{\rho{V}}$ and $G_{\rho{S}}$  are the isovector,  isovector-vector and isovector-scalar coupling constants, respectively. $\mathcal{L}_{IV}$ is also  chirally symmetric. Using the mean-field approximation,
\begin{eqnarray} \label{eq4}
 (\bar{\psi}A\psi)(\bar{\psi}B\psi)&=&(\bar{\psi}A\psi)<\bar{\psi}B\psi> +<\bar{\psi}A\psi>(\bar{\psi}B\psi) -<\bar{\psi}A\psi><\bar{\psi}B\psi>
\end{eqnarray}
the Lagrangian can be simplified to be
\begin{eqnarray} \label{eq5}
 \mathcal{L}&=&\mathcal{L}_{0}+\mathcal{L}_{IV}=\bar{\psi}[i\gamma_{\mu}{\partial^{\mu}}-m(\rho,\rho_S)- \gamma^0\Sigma(\rho,\rho_S,\rho_3)]\psi-U(\rho,\rho_S,\rho_3),
\end{eqnarray}
where $m$, $\Sigma$ and $U$ are defined as
\begin{eqnarray}
 m(\rho,\rho_S)&=&m_0-(G_S+G_{SV}\rho^2+G_{{\rho}S}\rho_3^2)\rho_S, \label{eqgap} \\
 \Sigma(\rho,\rho_S,\rho_3)&=&G_V\rho+G_{\rho}\rho_3\tau_3-G_{SV}\rho_{S}^{2}\rho-G_{{\rho}V}\rho_{3}^{2}\rho -G_{{\rho}V}\rho_{3}\rho^2\tau_3-G_{{\rho}S}\rho_{3}\rho_S^2\tau_3,\label{eq7}\\
 U(\rho,\rho_S,\rho_3)&=&\frac{1}{2}(G_S\rho_{S}^{2}-G_V\rho^2-G_\rho\rho_{3}^2+3G_{SV}\rho_{S}^{2}\rho^2 +3G_{{\rho}V}\rho_{3}^{2}\rho^2+3G_{{\rho}S}\rho_{3}^2\rho_S^2). \label{eq8}
\end{eqnarray}
Eq.(\ref{eqgap}) is the gap equation for the nucleon effective mass  in the NJL model. Here $\rho=<\bar{\psi}\gamma^0\psi>$ , $\rho_3=<\bar{\psi}\gamma^0\tau_3\psi>$ and $\rho_S=<\bar{\psi}\psi>$ are vector, isovector and  scalar densities, respectively.
From  the energy-momentum tensor, we may obtain  the following energy density and pressure
\begin{eqnarray} \label{eq11}
\epsilon&=&-\mathop{\sum}_{i=p,n}\nu_i\int_{p_{F_i}}^{\Lambda}\frac{d^3p}{(2\pi)^3}(p^2+m^2)^{1/2}+ \frac{G_V\rho^2}{2} +\frac{G_{\rho}\rho_{3}^2}{2}
+\frac{G_S\rho_{S}^2}{2}+\frac{G_{SV}\rho^2\rho_{S}^2}{2}\nonumber\\
&&- \frac{G_{\rho{V}}{\rho_3}^2\rho^2}{2}+\frac{G_{\rho{S}}{\rho_3}^2\rho_S^2}{2}+\epsilon_0,\\
P&=&-\mathop{\sum}_{i=p,n}\frac{\nu_i}{3}\int_{p_{F_i}}^{\Lambda}\frac{d^3k}{(2\pi)^3}\frac{k^2}{\sqrt{k^2+m^2}}+  \frac{G_{V}\rho^{2}}{2} +\frac{G_{\rho}\rho_{3}^{2}}{2}
 -\frac{G_S\rho_{S}^{2}}{2}-\frac{3G_{SV}\rho_{S}^{2}\rho^2}{2}\\
&&-\frac{3G_{\rho{V}}\rho_{3}^{2}\rho^2}{2}-\frac{3G_{\rho{S}}{\rho_3}^2\rho_S^2}{2}- \frac{2\Lambda^3\sqrt{\Lambda^2+m^2}}{3\pi^2}- \epsilon_0,
 \label{eq13}
 \end{eqnarray}
where   $\Lambda$ is the  momentum cutoff, and the $\epsilon_0$ is introduced to give the vanishing energy density of the vacuum state.
From the energy density, we can derive the symmetry energy as
\begin{eqnarray} \label{eq14}
E_{sym}(\rho)&=&\left.\frac{1}{2}\frac{\partial^2(\epsilon/\rho)}{\partial{\delta}^2}\right|_{\delta=0}=
\frac{p_{F}^2}{6E_F}+\frac{1}{2}G_{\rho}\rho -\frac{1}{2}G_{\rho{V}}\rho^3-\frac{1}{2}G_{\rho{S}}\rho_S^2\rho,
\end{eqnarray}
where $\delta=(\rho_n-\rho_p)/\rho$ is the isospin asymmetry parameter and $E_F=\sqrt{p_F^2+m^2}$.   The symmetry energy has a term linear in  $\rho^3$ due to the isovetor-vector interaction. The slope of the symmetry energy at saturation density is defined as
\begin{equation}
L=3\rho_0\left.\frac{\partial E_{sym}}{\partial\rho}\right|_{\rho_0}.
\end{equation}

\section{Results and discussions}

The present model has eight parameters: $\Lambda$, $m_0$, $G_S$, $G_V$, $G_{SV}$, $G_{\rho}$, $G_{\rho{V}}$, and $G_{{\rho}S}$. It was pointed out in Ref.~\cite{in12} that  $\Lambda > 0.6 GeV$ should be excluded, because otherwise the bare nucleon mass $m_0$ would be smaller than $3m_{0q}$, where $m_{0q}=(5\pm1)$ MeV ~\cite{k15} is the isospin-averaged current mass of light quarks.  Indeed, the cutoff larger than 600 MeV (with $m_0<3m_{0q}$) declines a monotonous decrease of the nucleon mass with the increase of density~\cite{in12}, thus disfavoring the characterization of  the in-medium chiral symmetry restoration. Here, the link between the bare nucleon mass and the current quark mass can be understood upon the constituent quark picture where the current quarks are released out after the chiral symmetry is restored.  Following Ref.~\cite{in12},  we  choose $\Lambda=400 MeV$ unless otherwise indicated. Note that the cutoff regularization can lead to an unphysical chiral condensate (also the scalar density $\rho_S$)  above the critical density corresponding to $p_F> \Lambda$. To avoid the unphysical chiral condensate at $p_F> \Lambda$, one may either  include a smooth cutoff function~\cite{jl1}, or set the relevant coupling constants   $G_S$, $G_{SV}$ and $G_{\rho S}$ to be zero~\cite{bra13}. In so doing, the nucleon effective mass will not fall below the bare mass $m_0$ at $P_F>\Lambda$, while we note these treatments  do not have significant effects on the asymmetric matter EOS because the nonzero scalar density of this model remains nearly vanishing at high densities.    Using Eq.(\ref{eqgap}) and $m_{\pi}^2 f_{\pi}^2$=$m_0{\rho}_{S}^{\rm{vac}}$, we obtain $m_0=41.3 MeV$ and $G_S=1.669$ $GeV\cdot fm^3$. This small bare nucleon mass interprets the consistency with the understanding that the mass acquisition arises dominantly from the non-perturbative vacuum. We note that a  different parametrization with very large bare mass is considered in a similar model~\cite{Pais16}. The saturation requirement, $(\epsilon/\rho)_{\rho=\rho_0}-m_N=-16MeV$ with $\rho_0=0.16fm^{-3}$ and $m_N$ being   the nucleon mass in the free space, gives $ G_V=1.581$ $GeV\cdot$$fm^3$ and $G_{SV}=2.054$  $GeV\cdot$$fm^9$.   The  coupling constants $G_{\rho V}$ and $G_{\rho S}$  are taken as  adjustable parameters to simulate different nuclear symmetry energies.  For vanishing  $G_{{\rho}V}$ and $G_{\rho S}$, we obtain $G_{\rho}$ to be 0.193 $GeV\cdot$ $fm^3$ by fitting the symmetry energy at saturation density to be 31.6 $MeV$~\cite{ba3}.

\begin{figure}[!htb]
\centering
\includegraphics[height=7.0cm,width=8cm]{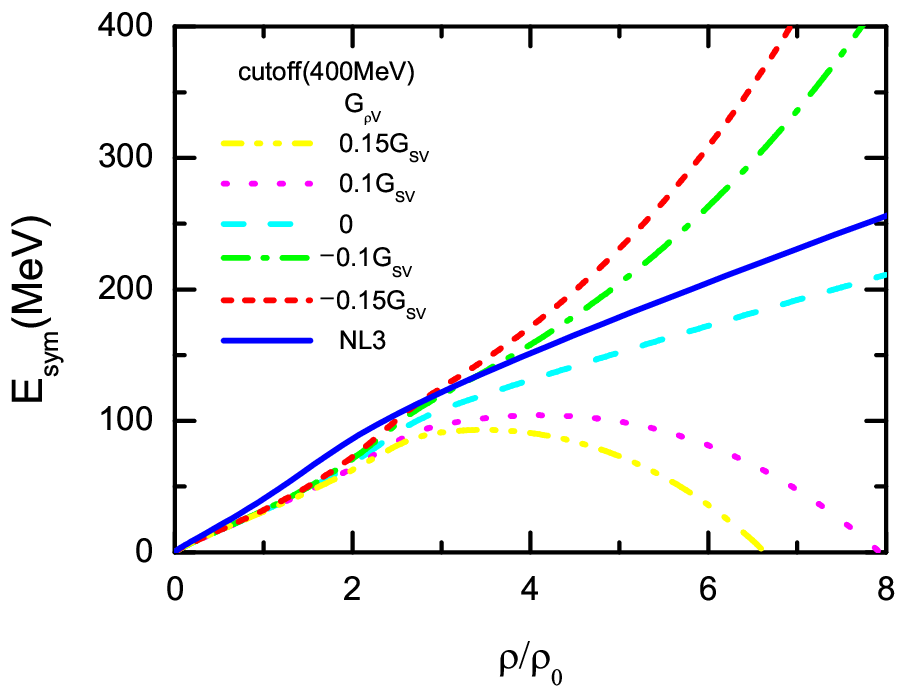}
\caption{ (Color online) The symmetry energy for  different $G_{\rho{V}}$ as a function of density. Here, $G_{\rho S}$ is set to be zero.  The symmetry energy with the RMF model NL3 is also depicted for comparison.}\label{fsym1}
\end{figure}
Fig.~\ref{fsym1} shows the symmetry energy  for different $G_{\rho{V}}$. For comparison, we also depict the symmetry energy with the nonlinear RMF model NL3~\cite{nl3}.  We can see that the symmetry energy without the isovector-vector interaction is softer  than  that with the NL3, while both evolve similarly with the density.  By adjusting the parameter  $G_{\rho{V}}$, we can simulate various density profiles of the symmetry energy that were reported in the model predictions~\cite{chen05,fu06} and data extractions~\cite{xiao16,zhao17,ru11}. Since the isovector-vector  interaction contributes the symmetry energy a term that is cubic in density, as seen in Eq.(\ref{eq14}), the modification to the symmetry energy is decisive at high densities.  The symmetry energy rises stiffly  for negative $G_{\rho{V}}$, while it  becomes super-soft till to below zero at high densities for positive $G_{\rho{V}}$.

Similar to the variation of the symmetry energy, the ratio of protons to neutrons turns out to be very sensitive to the isovector-vector couplings. This similarity lies in the fact that the difference between the proton and neutron chemical potentials, associated with the proton fraction,  is linear in the symmetry energy. For negative $G_{\rho V}$, the proton fraction increases with the increase of density, while for positive  $G_{\rho V}$ it first increases up to a maximum and then reduces with the increase of density. Corresponding to the super-soft symmetry energy with $G_{{\rho}V}=0.1G_{SV}$ and $0.15G_{SV}$, the proton fraction tends to disappear at high densities, which means that in the NS interior pure neutron matter arises~\cite{wi88,sz06}.  For  the vanishing proton fraction, the  $\rho^2\rho_3^2$ term becomes proportional to  $\rho^4$. This results in the dramatic reduction of the pressure   at high densities. The isovector-scalar interaction with the appropriate sign of $G_{\rho S}$ may produce some cancelation against the dramatic decrease of the pressure caused by the isovector-vector interaction. Such a cancelation is, however,  negligible because of the vanishing $\rho_S$ at high densities.  As a result, the EOS with the super-soft symmetry energy in the NJL model can not stabilize the NS.   The similar NS stability problem  was also found using the nonrelativistic models with the MDI interactions~\cite{we09}. While the over-reduced pressure was compensated by invoking the weakly interacting light U-boson~\cite{we09}, such a compensation would actually not help much in the present case because the  isovector-vector interaction reduces the pressure in a form linear in $\rho^4$. Therefore, the super-soft symmetry energy should eventually be excluded  in the NJL model.   We have noticed that the negative symmetry energies are disfavored by the stability arguments that belongs indeed to the positivity conditions on the second derivatives of the total energy and can be expressed in terms of Landau parameters~\cite{ba75,ma02,du12}.   We would, however, say that the present conclusion does not have to be universal to other approaches that account for high-order residual interactions.  For instance, in the presence of the super-soft symmetry energy,   the pressure of neutron star matter  may increase with the density in a non-relativistic microscopic calculation with the variational method~\cite{wi88}.

\begin{figure}[!htb]
\centering
\includegraphics[height=11.5cm,width=8cm]{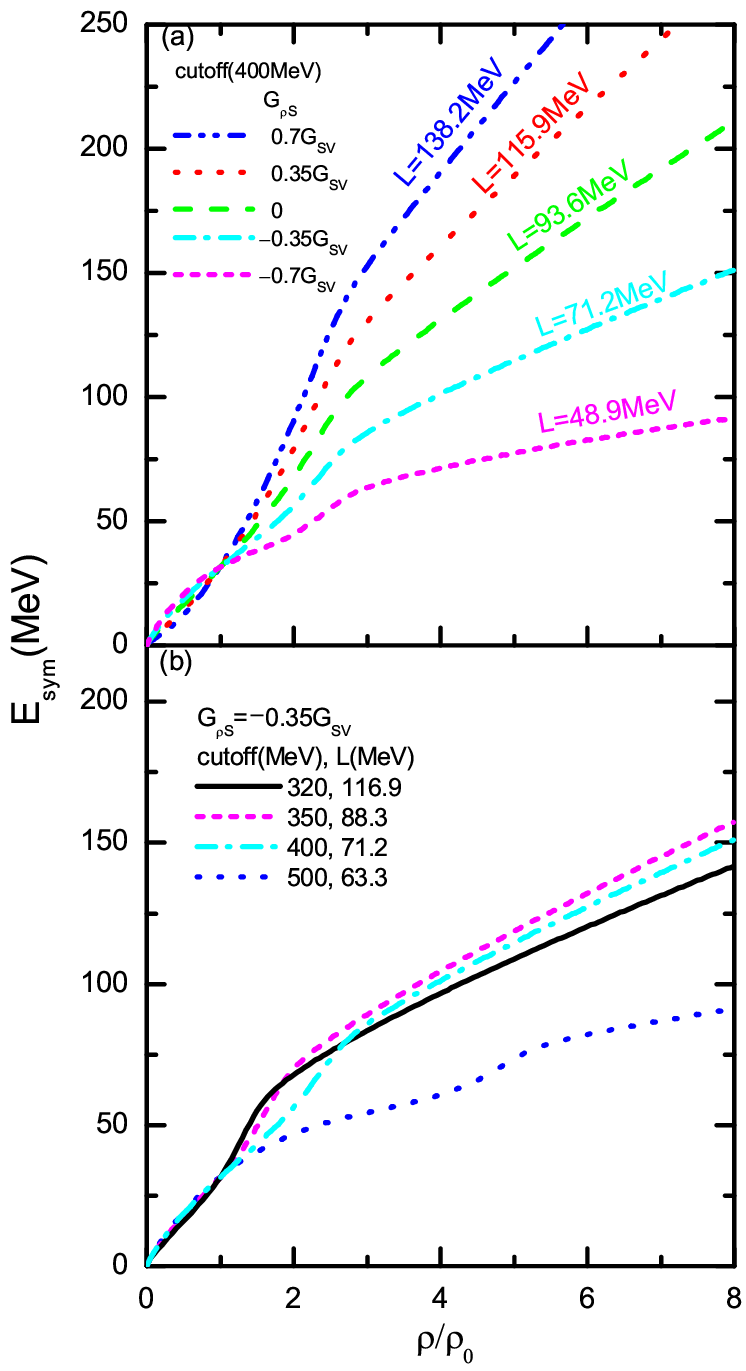}
\caption{ (Color online)The symmetry energy with various isovector-scalar couplings at $\Lambda=400 MeV$ (upper panel)  and  various cutoffs at  $G_{\rho S}=-0.35G_{SV}$ (lower panel). Here, $G_{SV}$ is different for various $\Lambda$, see Table~\ref{tb1}. }\label{fsym2}
\end{figure}

To further check the density dependence of the symmetry energy, we calculate the slope parameter of the symmetry energy at saturation density.  In the following calculation,  we neglect the  isovector-vector interaction due to the exclusion of the super-soft symmetry energy. Currently, the different extraction of the slope of the symmetry energy gives an average around $L\sim~40-60 MeV$~\cite{ba3,ch14,ts10,la13,wang15}. With $\Lambda=400MeV$, $L$ is 93.6$MeV$. To reduce the slope parameter, we can not simply adjust the cutoff or the coupling constant $G_\rho$. A feasible way is to invoke  the isovector-scalar interaction. Shown in the upper panel of Fig.~\ref{fsym2} is the symmetry energy with various isovector-scalar couplings. Here, the symmetry energy at saturation density is fixed to be 31.6 $MeV$ by adjusting the parameter $G_{\rho}$ for various $G_{{\rho}S}$, and $G_\rho$ is 0.006, 0.100, 0.286 and 0.379 $GeV\cdot fm^3$ for $G_{{\rho}S}/G_{SV}$=-0.70, -0.35, 0.35, and 0.70, respectively. We see that  the slope parameter can be reduced significantly by decreasing the $G_{\rho S}$. With $G_{\rho S}=-0.7G_{SV}$, the slope parameter is $48.9MeV$, being well within the average domain of extracted values. Note that the incompressibility is a constant ( $\kappa=296 MeV$) in this case and at densities 1.2-2.2$\rho_0$ the parametrization with $\Lambda=400MeV$   satisfies the constraints of the symmetric matter pressure  from the KaoS experiments~\cite{WG1}. A consequence of the parabolic approximation of the nuclear EOS is that pure neutron matter can be well specified by the density dependence of the symmetry energy and symmetric matter EOS that are both well constrained. Indeed, we find that the pressure of pure neutron matter with the parametrization of $\Lambda=400MeV$ and $G_{\rho S}=-0.7G_{SV}$ can satisfy the constraints from microscopic calculations based on chiral nucleon-nucleon and three-nucleon interactions~\cite{KH1}.

\begin{table*}[htp]
\caption{Parameter sets  for various cutoffs with $G_{\rho{S}}=-0.35G_{SV}$. $\rho_c$, evaluated by the relation $p_F=\Lambda$, is the critical density for chiral symmetry restoration. Listed in the last column is the incompressibility. Here, $G_\rho$ and $G_V$ are in unit of  $GeV\cdot$$fm^3$, and $G_{SV}$ and $G_{\rho{S}}$ are  in unit of $GeV\cdot$$fm^9$.\label{tb1}}
\begin{center}
\begin{tabular}{ c | c    |     c    |      c     |      c |      c|      c|      c|      c}
\hline\hline
  $\Lambda$(MeV) &$\rho_{c}/\rho_0$&   $G_S$  &      $m_0$(MeV)        &    $G_{SV}$     & $G_V$    &$G_{\rho}$  &$G_{\rho{S}}$ &$\kappa$ $(MeV)$\\ \hline
  $320$    &1.81      &    3.067            &        79.2            &       4.553          &  2.736        &  0.0848&-1.594&318\\
  $350$    &2.37      &    2.409            &        60.9            &       3.482          &  2.173        &  0.1095&-1.219&262\\
  $400$    &3.53      &    1.669            &        41.3            &       2.054          &  1.581        &  0.0996&-0.719&296\\
  $500$    &6.90      &    0.896            &        21.7            &       0.879          &  1.156        &  0.0058&-0.314&315\\
\hline
\hline
     \end{tabular} \end{center}
  \end{table*}

While we use the momentum cutoff $\Lambda=400MeV$ in above, it is now significant to examine how the results change with  the cutoff.  In the lower panel of Fig.~\ref{fsym2}, we display  the symmetry energy  with various  cutoffs at $G_{\rho S}=-0.35G_{SV}$. For different cutoffs, the parameter sets that maintain the saturation at $\rho_0=0.16fm^{-3}$ are tabulated in  Table.~\ref{tb1}.    As seen from the lower panel of Fig.~\ref{fsym2}, the symmetry energy with different cutoffs may be close at high densities. The reason for this to occur is that the $G_{\rho}$, determined by the symmetry energy at saturation density, is close  for different cutoffs, see Table~\ref{tb1}.  At high densities, the  term of $G_{\rho}$ dominates the symmetry energy (see Eq.(\ref{eq14}), since the nucleon mass and scalar density are small for the restoration of chiral symmetry.   It is interesting to see that the soft symmetry energy (at high densities) is not above the stiff one at lower densities, different from  those in the literature, e.g., see Ref.~\cite{chen05,ji07a}. We may attribute this to the behavior of the effective nucleon mass in the NJL model: there is a critical point because of the disappearance of the $\rho_S$ at $p_F=\Lambda$. With the increase of the cutoff, the critical density rises, and the similar tendency of the symmetry energy below and above saturation density still exists but fades away, while for the fixed cutoff the consistent softening of the symmetry energy at lower and high densities does not appear for various $G_{\rho S}$, as shown in the upper panel of Fig.~\ref{fsym2}. Nevertheless, a general softening of the symmetry energy at high densities are observed for various cases in both panels of  Fig.~\ref{fsym2}  because of the turning point concerning the restoration of  the chiral symmetry. This is rather remarkable because   the case of the large slope parameter of the symmetry energy   usually indicates  the stiff symmetry energy.

\begin{figure}[!htb]
\centering
\includegraphics[height=11.5cm,width=8cm]{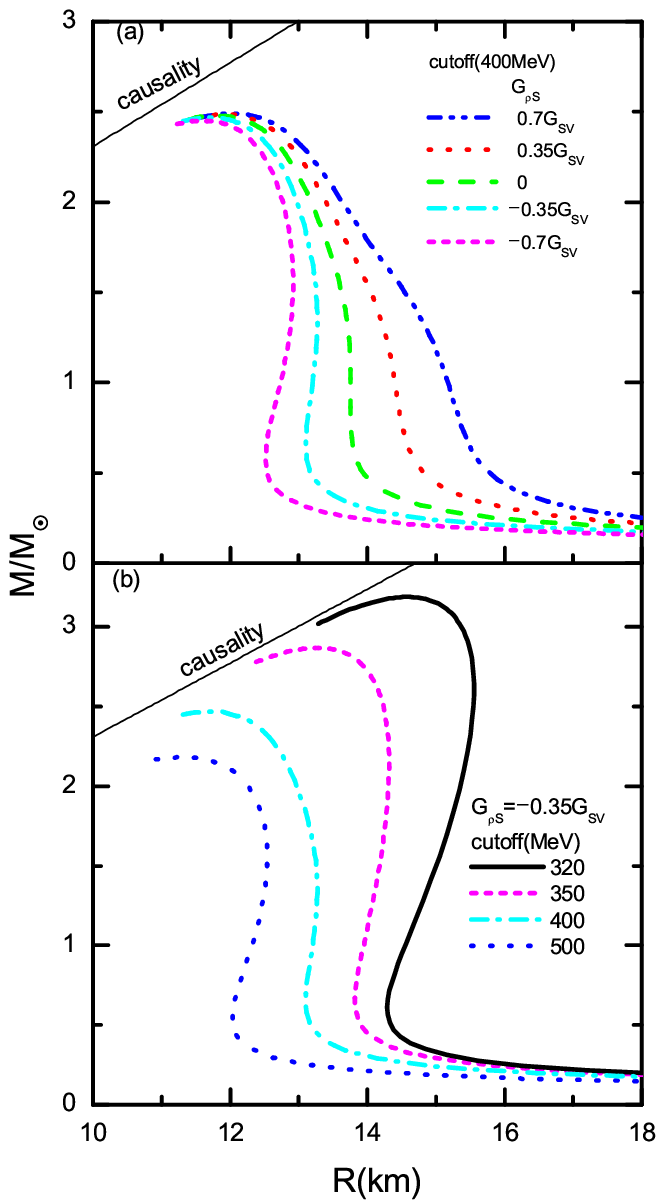}
\caption{ (Color online) The same as in Fig.~\ref{fsym2} but for NS mass-radius trajectories. }\label{fns2}
\end{figure}
\begin{table*}[htp]
\caption{Some specific NS properties with cases in Fig.3. $M_c$ is the NS mass by taking $\rho_c$ as its central density $\rho_{cen}$. The radius $R$ is in unit of km.\label{tb2}}
\begin{center}
\begin{tabular}{ c | c    |     c    |      c     |      c}
\hline\hline
  $\Lambda$(MeV) &$G_{\rho S}$ & ($\rho_{c}/\rho_0$, $M_c/M_{\odot}$, R )   &      ($\rho_{cen}/\rho_{0}$ , $M_{max}/M_{\odot}$, R )        &    R(1.4$M_{\odot}$)     \\ \hline
  $320$    &  -0.35$G_{SV}$      &    (1.81, 2.16, 15.4)         &         (3.38, 3.15, 14.5)            &       14.9          \\
  $350$    &-0.35$G_{SV}$        &     (2.37, 2.19, 14.3)        &       (4.18, 2.83, 13.2)            &       14.2         \\
  \hline
      &  0.7$G_{SV}$        &       (3.53, 2.18, 13.2)      &       (5.46, 2.43, 11.9)            &       14.8         \\
       &  0.35$G_{SV}$       &     (3.53, 2.18, 13.0)        &       ( 5.46 ,2.43, 11.8)            &       14.2         \\
  $400$    &    0                &       (3.53, 2.16, 12.9)      &       ( 5.48, 2.44, 11.7)           &       13.7         \\
       &  -0.35$G_{SV}$      &     (3.53, 2.15, 12.8)        &       (5.52, 2.44, 11.7)            &       13.3         \\
       &-0.7$G_{SV}$         &    (3.53, 2.13, 12.6)         &        (5.57, 2.43, 11.6)           &       13.0          \\
   \hline
  $500$    & -0.35$G_{SV}$       &     (6.90, 2.16, 11.0)        &       (6.06, 2.17, 11.3)           &       12.6         \\
\hline
\hline
     \end{tabular} \end{center}
  \end{table*}
Now, we turn to the  NS properties with the EOS obtained in the NJL model. The NS mass-radius relation  can be obtained by solving  the standard
Tolman-Oppenheimer-Volkoff (TOV) equation~\cite{Op39,Tol39}.
Here, we consider the simple compositions for NS matter: neutrons, protons, electrons and muons.
We adopt the EOS's obtained in this work at densities above half the saturation density, while we employ the standard low-density EOS~\cite{Ba71,Ii97} since at lower densities  NS matter transitions to inhomogeneous phase.  Shown in Fig.~\ref{fns2} are the NS mass-radius relations with the cases same as in Fig.\ref{fsym2}.  As seen in the upper panel of Fig.~\ref{fns2},  the NS maximum mass  does not change much by the isovector-scalar coupling,  since the latter does not have significant effects on the high-density EOS that dominates the NS maximum mass. While  the NS  radius is primarily determined by the slope of the symmetry energy in the density
range of 1 to $2\rho_0$~\cite{la07,la01},  the different $L$, shown in the upper panel of Fig.~\ref{fsym2}, can account for the large extent of different NS radii. The radius of a canonical NS without the isovector-scalar coupling  is about 13.7 km, locating at a reasonable position among various predictions ~\cite{ji07a,la07,li06} and extractions from recent observations~\cite{st16,bog13,pou14,hei14,latt14,oz15} ranging roughly from 10 to 15 km.   With decreasing the isovector-scalar coupling, the NS radius reduces accordingly. For instance, with $G_{\rho S}=-0.7G_{SV}$, the radius of the $1.4M_\odot$ NS is decreased to be 13.0 $km$.

Shown in the lower panel of Fig.~\ref{fns2} are the mass-radius relations for various cutoffs at  $G_{\rho S}=-0.35G_{SV}$. All cases  satisfy  the maximum mass constraint~\cite{pd14,ja15}.   We see that the NS maximum mass is significantly larger for smaller cutoffs due to  the stiffening of the high-density EOS. Since the scalar density that determines the nucleon effective mass almost vanishes beyond the critical density for chiral symmetry restoration, the vector term, denoted by the coupling $G_V$, then dictates the stiffness of the EOS at high densities. While  $G_V$ is larger  for smaller cutoffs, see Table~\ref{tb1},   the EOS beyond the critical density becomes stiffer with the decrease of the cutoff, resulting in larger NS maximum mass.
Corresponding to Fig.~\ref{fns2}, we tabulate some specific NS properties in Table~\ref{tb2}: the maximum mass, corresponding radius and central density, and the radius of a $1.4M_\odot$ star. As a comparison, we also tabulate the results with the central density being the chiral symmetry restoration density $\rho_c$. We can find that the asymmetric matter EOS beyond $\rho_c$ can have a significant contribution, resulting dominantly from the vector term,  to the NS maximum mass, especially for small cutoffs. Note that such a contribution is almost independent of the situation whether or not we have removed the unphysical chiral condensate beyond $\rho_c$ using the methods in Refs.~\cite{jl1,bra13}.
It is worthy to point out that the pressure of symmetric matter with  $\Lambda=500MeV$ can well satisfy the constraints from collective
flow data in heavy-ion collisions~\cite{PD1}, and the pressure with small cutoffs may surpass the constraints. The pressure of symmetric matter with  $\Lambda=400MeV$ surpasses the flow data constraints beyond 3 $\rho_0$, but is not far above the upper limit. While the pressure of pure neutron matter with various symmetry energies is within or close to the region allowed by the flow data especially at high densities, the predicted NS maximum mass with $\Lambda=400 MeV$ is rather acceptable.   The parametrizations with $\Lambda=350-400MeV$ can well fit the constraints from the microscopic calculations for pure neutron matter~\cite{KH1}, while the parametrization with smaller cutoff, e.g., $350MeV$, may easily surpass the flow data constraints and produce a large NS maximum mass.  Similar to the case in the upper panel of Fig.~\ref{fns2},  the   NS radii with various cutoffs are  associated with the slope parameter $L$.  Here, the $L$ is 116.9, 88.3, 71.2, and 63.3 $MeV$ with the cutoff 320, 350, 400 and 500 $MeV$, respectively. This is roughly corresponding to different NS radii, as shown in the lower panel of Fig.~\ref{fns2}. We should, however, note that different cutoffs can result in the difference in  properties  of  symmetric matter that also contributes to the large separation in NS radii. For instance, rather different incompressibility at saturation density arises for various cutoffs,  as seen in Table~\ref{tb1}.   Our investigation indicates that a  combination of favorably large cutoffs and (negative) isovector-scalar couplings in the saturated NJL model can result in relatively small NS radii which are consistent with those extracted from recent measurements~\cite{st16,bog13,pou14,hei14,latt14,oz15}. We may reasonably require the positive  $G_\rho$ for any $G_{\rho S}$ to fit the symmetry energy at saturation density. Within the cutoff range of 320-500MeV  for non-positive $G_{\rho S}$, we  estimate the radius region of the $1.4M_\odot$ NS to be around 12.6-14.9km, see Table~\ref{tb2}.

\section{Summary}

In this work, we adopt the saturated NJL model that respects the chiral symmetry  to study the density dependence of the symmetry energy and its consequence in NS's. While the super-soft symmetry energy can not be  produced by the usual RMF models, we find that  a chiral isovector-vector interaction can be responsible for the super-soft symmetry energy, though the latter should eventually be ruled out by the NS stability. With the inclusion of the isovector-scalar interaction, a general softening of the symmetry energy at high densities is found even for the large slope parameter of the symmetry energy at saturation density because of the restoration of  chiral symmetry.  We have also examined the dependence of the symmetry energy on the momentum cutoff of the NJL model. The rise of the cutoff in a reasonable region reduces the slope of the symmetry energy at saturation density. For smaller cutoffs, the symmetry energy in the  NJL model may display  consistent stiffness or softness on the both sides of the saturation density. Finally, using the NJL EOS's, we have investigated the NS  mass-radius relations. The NS maximum mass obtained with various isovector-scalar couplings and momentum cutoffs is well above the $2M_\odot$. The relatively small NS radius can be obtained with suitable combination of reasonable cutoffs and isovector-scalar couplings, and the obtained NS radii  well meet the present limits extracted from recent measurements.

\section*{ACKNOWLEDGMENT}
We thank Profs. Lie-Wen Chen and Hong-Shi Zong for helpful discussions.
The work was supported in part by the National Natural Science
Foundation of China under Grant No. 11275048 and the China Jiangsu
Provincial Natural Science Foundation under Grant No. BK20131286.

\end{document}